\documentclass[aps,prl,10pt,amsmath,twocolumn,superscriptaddress,amssymb,showpacs]{revtex4-1} 
\usepackage{graphicx}
\usepackage{color}
\usepackage[tight]{units}
\usepackage[]{nicefrac}
\usepackage{mathptmx}      
\usepackage{amssymb}
\newcommand{\BiTe}{Bi$_{2}$Te$_{3}$~}
\newcommand{\SbTe}{Sb$_{2}$Te$_{3}$~}
\newcommand{\BiTex}{Bi$_{2}$Te$_{3}$}
\newcommand{\SbTex}{Sb$_{2}$Te$_{3}$}
\newcommand{\SB}{Bi$_{2}$Te$_{3}$/Sb$_{2}$Te$_{3}$}

\newcommand{\SBxSL}{(Bi$_2$Te$_3)_{x}$/(Sb$_2$Te$_3)_{1-x}$ SL~}

\newcommand{\SBSLsx}{Bi$_{2}$Te$_{3}$/Sb$_{2}$Te$_{3}$ SLs}

\newcommand{\Vline}{Ref.~\cite{Venkatasubramanian:2001p114}~}

\newcommand{\f}[1]{Fig.~\ref{fig:#1}}

\begin{document}

\title{Lorenz function of Bi$_{2}$Te$_{3}$/Sb$_{2}$Te$_{3}$ superlattices}

\author{N. F. Hinsche}
\email{nicki.hinsche@physik.uni-halle.de}
\affiliation{Institut f\"{u}r Physik, Martin-Luther-Universit\"{a}t Halle-Wittenberg, DE-06099 Halle, Germany}
\author{I. Mertig}
\affiliation{Institut f\"{u}r Physik, Martin-Luther-Universit\"{a}t Halle-Wittenberg, DE-06099 Halle, Germany}
\affiliation{Max-Planck-Institut f\"{u}r Mikrostrukturphysik, Weinberg 2, DE-06120 Halle, Germany}
\author{P. Zahn}
\affiliation{Helmholtz-Zentrum Dresden-Rossendorf, P.O.Box 51 01 19, DE-01314 Dresden, Germany}
%
\date{\today}


\begin{abstract}
Combining first principles density functional theory and semi-classical Boltzmann transport, 
the anisotropic Lorenz function was studied for thermoelectric Bi$_{2}$Te$_{3}$/Sb$_{2}$Te$_{3}$ superlattices 
and their bulk constituents. It was found that already for the bulk materials \BiTe and \SbTex, 
the Lorenz function is not a pellucid function on charge carrier concentration and temperature. 
For electron-doped Bi$_{2}$Te$_{3}$/Sb$_{2}$Te$_{3}$ superlattices large oscillatory deviations for 
the Lorenz function from the metallic limit were found even at high charge carrier concentrations. 
The latter can be referred to quantum well effects, which occur at distinct superlattice periods.
\end{abstract}

\maketitle


\section{Introduction}
\label{intro}
For many decades thermoelectric (TE) energy conversion successfully enabled self-supporting energy 
devices for outer-space missions or integrated electronics 
\cite{Sales:2002p6580,Tritt:2006p15694}. However, bad conversion efficiency 
prohibited thermoelectrics the break-through as an alternative energy source. 
The conversion performance of a TE material is quantified by the 
figure of merit (FOM) 
\begin{equation}
ZT=\frac{\sigma S^{2}}{\kappa_{el} + \kappa_{ph}} T = \frac{ S^{2}}{L + \frac{\kappa_{ph}}{\sigma T}},
\label{ZT}
\end{equation}
where $\sigma$ is the electrical conductivity, $S$ the thermopower, $\kappa_{el}=L \sigma T$  and 
$\kappa_{ph}$ are the electronic and lattice contribution to the thermal conductivity, respectively. 
L denotes the Lorenz function, which becomes the 
Lorenz number L$_0=\frac{(\pi k_{b})^2}{3e^2}$ in the highly degenerate, metallic limit. 

In recent years nano-structuring concepts \cite{Bottner:2006p2812,Nolas:1999p15771} 
enable higher values for $ZT$ by increasing the 
numerator, called power factor $PF=\sigma S^{2}$, or decreasing 
the denominator of Eq.~\ref{ZT}. 
The latter is obtained by phonon-blocking at superlattice (SL) interfaces or grain boundaries  \cite{BorcaTasciuc:2000p15132,Lee:1997p1545,Pernot:2010p14944,Venkatasubramanian:2000p7305} 
and leads to a reduced lattice thermal conductivity $\kappa_{ph}$. 
Here, the Lorenz function is particularly important for thermoelectrics, 
providing a measure to separate the electronic and lattice contribution 
to the thermal conductivity \cite{Uher:1974p15736}. 
Deviations $L\neq L_0$ already occur in the degenerate limit for simple metals, semi-metals 
and semi-conductors \cite{Kumar:1993p15734}. Hence, assuming incorrect values for the Lorenz number 
leads to incorrect values for $\kappa_{el}$ and $\kappa_{ph}$ and can even sum up to non-physically 
negative values for $\kappa_{ph}$ \cite{Sharp:2001p15840}. To the best of our knowledge, 
investigations on the Lorenz function of thermoelectric SLs on an \textit{ab initio} level are missing so far. 

\begin{figure}
\centering
  \includegraphics[width=0.48\textwidth]{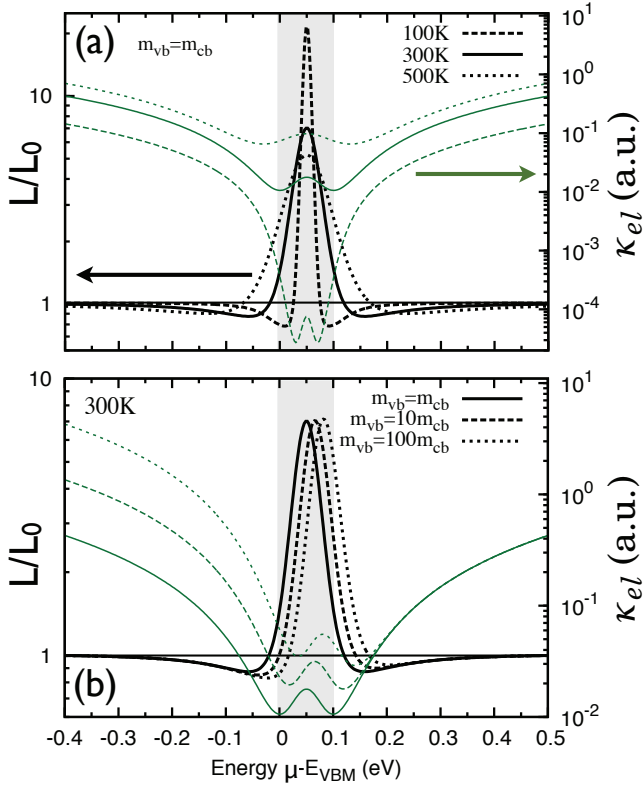}
\caption{Lorenz function L (thick black lines, ref. to left scale) and 
electronic contribution $\kappa_{el}$ to the total thermal conductivity (thin green lines, ref. to the right scale) 
in dependence on position of the chemical potential $\mu$ within a spherical two band 
model. Results are shown for (a) fixed effective masses $m_{vb}=m_{cb}$ and varying temperatures and 
(b) fixed temperature $T=\unit[300]{K}$ and varying effective masses. The band gap is fixed to $E_{g}=\unit[0.1]{eV}$ 
(gray shaded areas) and the Lorenz function is related 
to the metallic limit L$_0=\unit[2.44\times 10^{-8}]{W \Omega/K^{2}}$.}
\label{fig:1}
\end{figure}

In the present work we analyse the anisotropic Lorenz function for 
Bi$_{2}$Te$_{3}$/Sb$_{2}$Te$_{3}$ SLs, as well as for the bulk constituents. 
The two telluride single crystals and the composed p-type SL 
show highest values for bulk and nano-structured TE so far \cite{Venkatasubramanian:2001p114}. 
On the basis of \textit{ab initio} 
density functional theory (DFT) and semi-classical Boltzmann transport equations (BTE) the 
Lorenz function is in particular studied for different charge carrier concentrations and SL periods 
at room-temperature.

\section{Methodology}
\label{method}

For both \BiTe and \SbTex, as well as for the composed SLs, 
we used the experimental lattice parameters and relaxed atomic positions \cite{Landolt} 
as provided for the hexagonal \BiTe crystal structure. The 15 atomic layers per unit cell are 
composed out of three quintuples Te$_1$-Bi-Te$_2$-Bi-Te$_1$. 
The hexagonal lattice parameters are equally chosen to be  
${a^{hex}_{BiTe}}=4.384${\AA} and $c^{hex}_{BiTe}=30.487${\AA} for 
\BiTex, \SbTe and the SLs respectively. Preceding studies revealed 
that a larger in-plane lattice constant, e.g. ${a^{hex}_{BiTe}}>{a^{hex}_{SbTe}}$, 
is favourable for an enhanced cross-plane TE 
transport \cite{Hinsche:2011p15707,Yavorsky:2011p15466}. 
To introduce SLs with different SL periods we subsequently substitute the Bi sites 
by Sb, starting with six Bi sites in hexagonal bulk \BiTex. Substituting four atomic layers 
of Bi with Sb leads to a \SBxSL with $x=\frac{2}{6}$, that is one quintuple \BiTe and two quintuple \SbTex. 
The latter case coincides with a (10\AA/20\AA)-(\BiTex/\SbTex) SL 
in the experimental notation of \Vline.

Semi-classical BTE were extensively used in the past to calculate TE transport 
properties \cite{MZiman:1960p6024,Mertig:1999p12776} and offer a high reliability for 
narrow-gap semi-conductors in a broad doping and temperature range 
\cite{Hinsche:2011p15707,Park:2010p11006,Chaput:2005p1405,Huang:2008p559}. 
Within the relaxation time approximation (RTA) the transport distribution 
function (TDF) $\mathcal{L}_{\perp, \|}^{(0)}(\mu, 0)$~\cite{Mahan:1996p508} and with this the 
generalized conductance moments $\mathcal{L}_{\perp, \|}^{(n)}(\mu, T)$ are defined as 
\begin{eqnarray}
& \mathcal{L}_{\perp, \|}^{(n)}(\mu, T)= \nonumber \\
&\frac{\tau}{(2\pi)^3} \sum \limits_{\nu} \int\ d^3k \left( v^{\nu}_{k,(\perp, \|)}\right)^2 (E^{\nu}_k-\mu)^{n}\left( -\frac{\partial f_{(\mu,T)}}{\partial E} \right)_{E=E^{\nu}_k} \nonumber .
\\
\label{Tcoeff}
\end{eqnarray} 
$f_{(\mu,T)}$ is the \textsc{Fermi-Dirac}-distribution and $v^{\nu}_{k,(\|)}$, $v^{\nu}_{k,(\perp)}$ 
denote the group velocities in the directions in the 
hexagonal basal plane and perpendicular to it, respectively. Within here the group velocities 
were obtained as derivatives along 
the lines of the Bl\"ochl mesh in the whole Brillouin zone (BZ)~\cite{Yavorsky:2011p15466}. 
The band structure $E_k^{\nu}$ of band $\nu$ was obtained by accurate first principles density functional theory calculations (DFT), 
as implemented in the fully relativistic screened Korringa-Kohn-Rostoker Greens-function method (KKR) \cite{Gradhand:2009p7460}. 
Within this approach the \textsc{Dirac}-equation is solved self-consistently and with that spin-orbit-coupling (SOC) is included. 
Exchange and correlation effects were accounted for by the local density 
approximation (LDA) parametrized by Vosko, Wilk, and Nusair \cite{Vosko1980}. Detailed studies on the electronic 
structure, the thermoelectric transport and challenges in the numerical determination 
of the group velocities of Bi$_{2}$Te$_{3}$, Sb$_{2}$Te$_{3}$ and their SLs have been published before 
\cite{Hinsche:2011p15707,Yavorsky:2011p15466,Zahn:2011p15523,ArxivSL}. 
For convenience, the relaxation time $\tau$ was chosen as $\unit[10]{fs}$ for the considered systems.

Straight forward, the temperature- and doping-dependent 
electrical conductivity $\sigma$ and thermopower $S$ 
in the in- and cross-plane directions are defined as
\begin{eqnarray}
\sigma_{_{\perp, \|}}=e^2 \mathcal{L}_{\perp, \|}^{(0)}(\mu, T) \qquad 
S_{_{\perp, \|}}=\frac{1} {eT} \frac{\mathcal{L}_{\perp, \|}^{(1)}(\mu,T)} {\mathcal{L}_{\perp, \|}^{(0)}(\mu,T)}
\label{Seeb},
\end{eqnarray}
and the electronic part to the total thermal conductivity accounts to
\begin{equation}
\kappa_{el}{_{\perp, \|}}=\frac{1}{T}(\mathcal{L}_{\perp, \|}^{(2)}(\mu,T)-\frac{\left(\mathcal{L}_{\perp, \|}^{(1)}(\mu,T)\right)^2}{\mathcal{L}_{\perp, \|}^{(0)}(\mu,T)}) \, .
\label{kel}
\end{equation}
The second term in eq.~\ref{kel} introduces corrections due to the Peltier heat flow that can occur when 
bipolar conduction takes place \cite{Goldsmid:1965p15735,Tritt:2004p15755}. Using Eqs.~\ref{Seeb} and \ref{kel} and 
the abbreviation $\kappa^{0}=\frac{1}{T}\mathcal{L}_{\perp, \|}^{(2)}(\mu,T)$~\cite{Mahan:1996p508}, 
we find the Lorenz function as 

\begin{equation}
L_{\perp, \|}=\frac{\kappa^{0}}{\sigma_{_{\perp, \|}} T}-S_{_{\perp, \|}}^{2}.
\label{Lorenz}
\end{equation}
Eq.~\ref{Lorenz} clearly shows that in the low temperature regime L consists of a constant 
term and a negative term of order $T^2$.

\section{Results}
\label{results}

To introduce our discussions, in \f{1} the Lorenz function L and 
the corresponding electronic thermal conductivity 
$\kappa_{el}$ in dependence on the chemical potential $\mu$ are shown 
for a spherical two band model (SBM). Varying temperatures and $m_{cb}=m_{vb}$ (cf. \f{1}(a)) and 
different effective masses ratio $\nicefrac{m_{cb}}{m_{vb}}$ and fixed temperature $T=\unit[300]{K}$ (cf. \f{1}(b)) are assumed. 
Within here $m_{cb}$ and $m_{vb}$ 
are the isotropic effective masses of the conduction band (CB) and valence band (VB), respectively. 
Setting the valence band maximum to zero and $E_g$ the band gap size, 
the TDF scales as $\mathcal{L}_{VB}^{(0)}(\mu, 0) \sim \sqrt{m_{vb}}(-\mu)^{3/2}$ and 
$\mathcal{L}_{VB}^{(0)}(\mu, 0) \sim \sqrt{m_{cb}}(\mu-E_{g})^{3/2}$ for the VB and CB, respectively. 
From Eqs.~\ref{kel} and \ref{Lorenz} it is obvious that within a SBM deviations for L and $\kappa_{el}$ 
from the metallic limit will merely occur near the band gap, where the thermopower S changes significantly. 
Near the band edges S increases 
approximately as $S \sim \frac{-1}{\mu T}$. Thus L, as well as $\kappa_{el}$ minimize and 
the minimum decreases with decreasing temperature, while shifting towards the middle of the gap (cf. \f{1}(a)). 
At $T=\unit[100]{K}$ $\nicefrac{L}{L_0}\sim 0.8$ at the band edges. 
In the intrinsic regime $\nicefrac{L}{L_0}$ and $\kappa_{el}$ increase, as the thermopower and electrical 
conductivity are reduced due to bipolar contributions. 
Figuratively speaking, the additional contribution arises from the fact that electron and holes can 
move together in the same direction, transporting energy but not carrying 
any net charge~\cite{Goldsmid:1965p15735}. 
According to Goldsmid~\cite{Goldsmid:1956p15499} and Price~\cite{Price:1956p15839} the deviation of 
the Lorenz number from the metallic limit in the intrinsic regime holds to some extent 
$\nicefrac{L}{L_0}=1+\frac{1}{2}\frac{m_{cb}m_{vb}}{(m_{cb}+m_{vb})^2} \left( \nicefrac{E_g}{k_B T+4} \right)^{2}$. 
Therefore, assuming a fixed charge carrier concentration, $\nicefrac{L}{L_0}$ achieves very large values 
at small temperatures and/or large band gaps. Assuming the above approaches~\cite{Goldsmid:1956p15499,Price:1956p15839}, 
together with $m_{cb}=m_{vb}$ and $E_{g}=\unit[0.1]{eV}$ one achieves $\nicefrac{L}{L_0} \sim 9$ 
at room temperature for $\mu$ located deep in the gap. 
If $m_{vb} > m_{cb}$, as shown in \f{1}(b), the intrinsic regime $N_n=N_p$ and with that the maximal value of 
$\nicefrac{L}{L_0}$ and $\kappa_{el}$ at bipolar conduction shifts towards the CBM. 
With increasing $m_{vb}$ and hence due to the enhanced electrical conductivity $\sigma$ in the VB it 
is obvious, that $\kappa_{el}$ under hole doping will increase, too. 

\begin{figure}
\centering
  \includegraphics[width=0.48\textwidth]{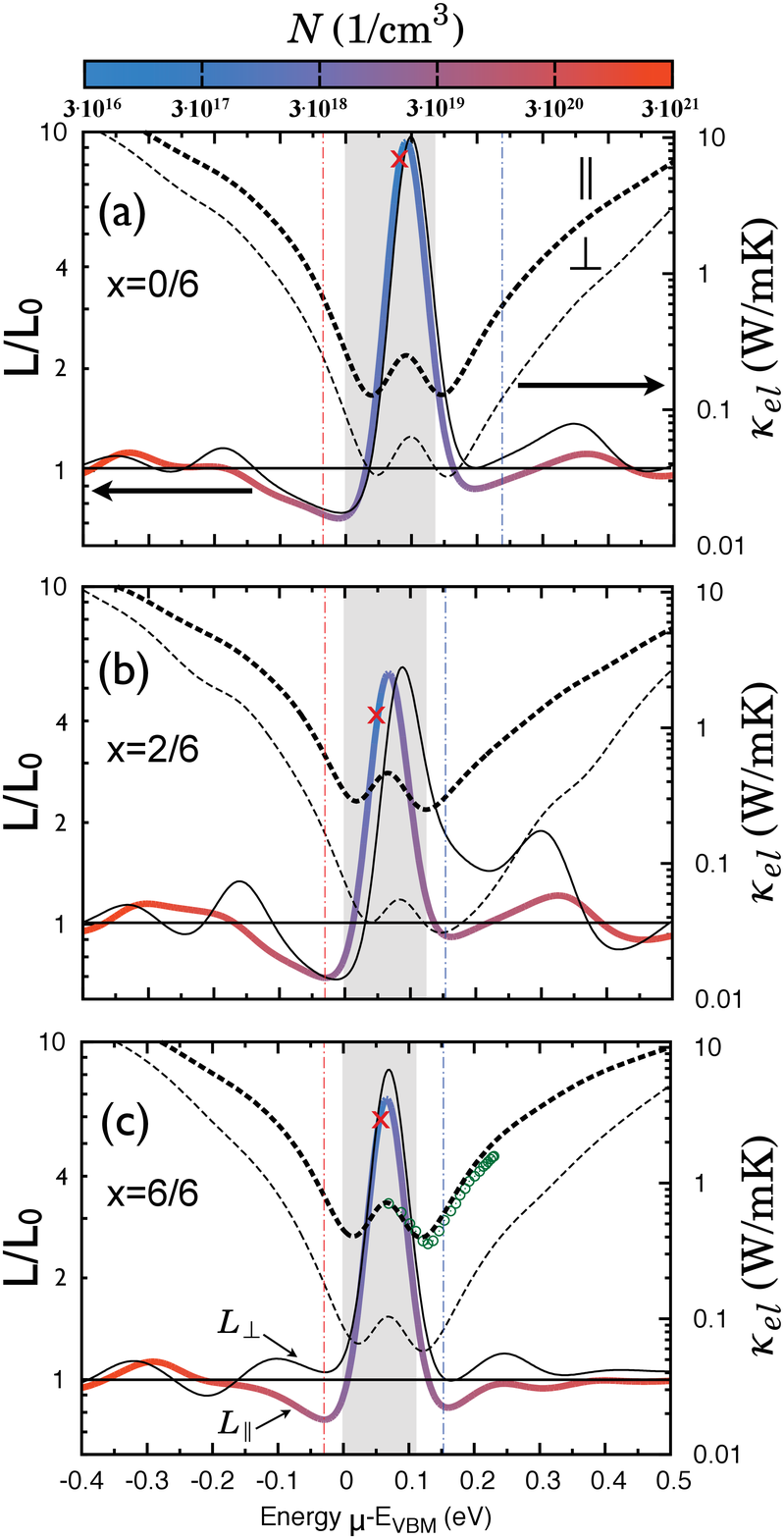}
\caption{Lorenz function L (solid lines, ref. to left scale) and 
electronic contribution $\kappa_{el}$ to the total thermal conductivity (dashed lines, ref. to right scale) 
in dependence on position of the chemical potential $\mu$ for 
(a) bulk Sb$_2$Te$_3$ (b) (Bi$_2$Te$_3)_{x}/($Sb$_2$Te$_3)_{1-x}$ SL at $x=\frac{2}{6}$ 
and (c) bulk Bi$_2$Te$_3$. The in-plane (thick lines) 
and cross-plane (thin lines) transport directions are compared. The Lorenz function is related to the 
metallic limit L$_0=\unit[2.44\times 10^{-8}]{W \Omega/K^{2}}$. 
Plotted on to the graph of the Lorenz function 
in the in-plane direction is a color code referring to the charge carrier concentration. 
The red cross emphasizes the change from n to p doping. 
The temperature was fixed to $\unit[300]{K}$. Thin vertical dash-dotted lines 
emphasize the position of the chemical potential for a 
charge carrier concentration of $N = \unit[3\times 10^{19}]{cm^{-3}}$ under 
p and n doping (red and blue color). The grey shaded areas show the band gap. 
Green open circles in (c) show experimental results from Ref.~\cite{Goldsmid:1965p15735} 
for $\kappa_{el,\|}$ for an n-type \BiTe single crystal.}
\label{fig:2}
\end{figure}

\f{2} presents first principle calculations for the Lorenz function L and the 
related electronic part $\kappa_{el}$ of the thermal conductivity. 
The dependence on the charge carrier concentration for (a) \SbTex, (b) a
(Bi$_2$Te$_3)_{x}/($Sb$_2$Te$_3)_{1-x}$ SL at $x=\nicefrac{2}{6}$ and (c) \BiTex, 
is shown, respectively. Due to the high conductivity anisotropy $\nicefrac{\sigma_{\|}}{\sigma_{\perp}}>1$ 
for all of the considered systems\cite{Hinsche:2011p15707,ArxivSL}, $\kappa_{el,\perp}$ 
is strongly suppressed compared to $\kappa_{el,\|}$, too. Furthermore, it is obvious 
that the maximal peak of the Lorenz function is shifted towards the CBM, latter stemming from 
a larger density of states at the VBM and a higher absolute hole electrical conductivity. Maximal numbers 
for the Lorenz function $\nicefrac{L}{L_0}$ in the intrinsic regime were found to be between 
6 and 10 for the considered systems, showing only a slight directional anisotropy. 
For \SbTe (cf. \f{2}(a)) the Lorenz function exhibits only minor anisotropies $\nicefrac{L_{\|}}{L_{\perp}}$ 
in a wide doping range, while stating $L_{\perp} \sim 1.15 L_{\|}$ at increased electron doping. 
Reduction of $\nicefrac{L_{\|}}{L_{\perp}}$ due to bipolar diffusion effects is more apparent at 
hole doping compared to electron doping, here showing in-plane $\nicefrac{L}{L_0} \sim 0.75$ and 
$\nicefrac{L}{L_0} \sim 0.92$ at an electron and hole doping of $N = \unit[3\times 10^{19}]{cm^{-3}}$, respectively. 
For bulk \BiTe the picture is comparable. However, in the thermoelectric most interesting range, 
about $\unit[200]{meV}$ around the band edges, $\nicefrac{L_{\|}}{L_{\perp}}$ is always less than unity, 
comparable to previous publications~\cite{Huang:2008p559}. 
Furthermore $L_{\perp}$ is larger than the metallic limit $L_0$. Very often values of $L_{\perp} \sim 0.5-0.6$~\cite{Beyer:2002p15267,Venkatasubramanian:2001p114} 
are assumed for the experimental determination of $\kappa_{el,\perp}$ in \SBSLsx. In turn, 
this most probably leads to an underestimation of the electrical contribution to total thermal conductivity 
in cross-plane direction. For bulk \BiTe experimental values~\cite{Goldsmid:1965p15735} 
for the in-plane part $\kappa_{el,\|}$ are available as a reference in \f{2}(c) (green, open circles). 
We find very good accordance to our calculations in the intrinsic range, 
while our results slightly overestimate $\kappa_{el,\|}$ in the extrinsic regime. 

Strong deviations for the Lorenz function from the bulk limit could be found for an electron conducting \SBxSL 
at $x=\nicefrac{2}{6}$, i.e. (10\AA/20\AA)-(\BiTex/\SbTex). 
In a current publication~\cite{ArxivSL} we showed, that strong quantum well effects (QWE) in the CB of the SLs lead to 
an enhanced electrical conductivity anisotropy, The latter was most pro\-noun\-ced for the SL at $x=\nicefrac{2}{6}$ 
showing $\nicefrac{\sigma_{\|}}{\sigma_{\perp}} \sim 20$ at electron doping of $N = \unit[3\times 10^{19}]{cm^{-3}}$. 
Caused by the QWE, the cross-plane electrical conductivity $\sigma_{\perp}$ is drastically suppressed, 
and hence $L_{\perp}$ remarkably enhanced. 
The cross-plane Lorenz function obtains rather large values between $\nicefrac{L}{L_0} \sim 1.5-1.8$ 
at extrinsic carrier concentrations of about $N = \unit[3-30\times 10^{19}]{cm^{-3}}$ for hole and electron 
doping, respectively. Additionally, 
oscillations of $\nicefrac{L}{L_0}$ with varying doping are found, which are much more pronounced than 
in the bulk materials. Both effects have been proposed within a 1-dimensional model for 
thermoelectric SLs before~\cite{Bian:2007p15769}. As expected, in the extrinsic region, 
at increasing charge carrier concentration, $L$ saturates gradually towards the metallic limit $L_0$ 
and the thermal conductivity rises with electrical conductivity. 
  
\begin{figure}
\centering
  \includegraphics[width=0.48\textwidth]{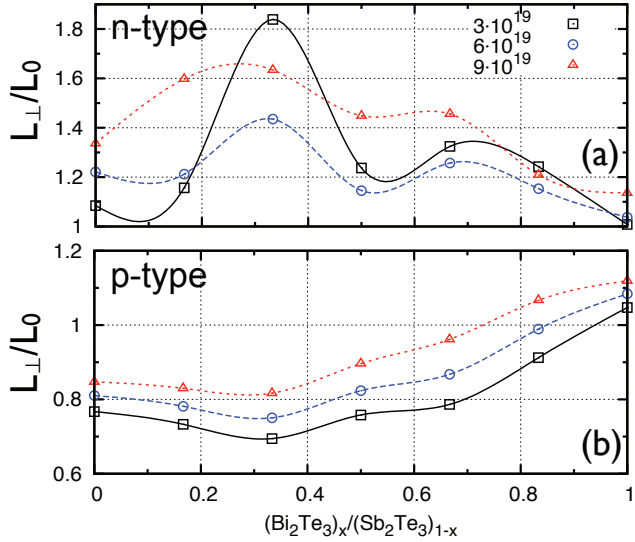}
\caption{Cross-plane component of the Lorenz function L$_{\perp}$ 
for (Bi$_2$Te$_3)_{x}/($Sb$_2$Te$_3)_{1-x}$ superlattices in dependence 
on the superlattice period. The temperature is fixed to $\unit[300]{K}$ and results for three 
different charge carrier concentrations (in units of $\unit[]{cm^{-3}}$)
are compared. (a) refers to electron doping, while (b) refers to hole doping. 
The Lorenz function is related to the metallic limit L$_0=\unit[2.44\times 10^{-8}]{W \Omega/K^{2}}$. 
Lines are guides to the eye.}
\label{fig:3}
\end{figure}

To support our findings, in \f{3} the cross-plane Lorenz function $L_{\perp}$ at different 
SL periods is shown. The influence of varying (a) electron and (b) hole doping is given, respectively. 
Under hole doping, due to the vanishing band-offset at the VBM~\cite{ArxivSL}, the Lorenz function behaves as smooth interpolation 
between the bulk limits (cf. \f{3}(b)). At lower charge carrier concentrations $\nicefrac{L_{\perp}}{L_0}$ 
is more suppressed due to a stronger impact of the bipolar diffusion. 
Under varying electron doping we find $\nicefrac{L_{\perp}}{L_0}$ being remarkably enhanced for 
SL periods of $x=\nicefrac{2}{6}$ and $x=\nicefrac{4}{6}$, respectively. For those SL periods 
large suppressions of the cross-plane electrical conductivity $\sigma_{\perp}$ were found, too. 
At $N = \unit[3\times 10^{19}]{cm^{-3}}$ anisotropies as large as  $\nicefrac{\sigma_{\|}}{\sigma_{\perp}} \sim 20$ and 
$\nicefrac{\sigma_{\|}}{\sigma_{\perp}} \sim 14$ for SL periods of $x=\nicefrac{2}{6}$ and $x=\nicefrac{4}{6}$ 
are reported, respectively \cite{ArxivSL}. We state that due to quantum confinement effects 
in the electron-conducting SLs unexpected deviations of the Lorenz function from $L_0$ 
can occur also at higher values of doping, which is counterintuitive. 
The latter could lead to wrong estimations for the electronic part of the thermal 
conductivity $\kappa_{el}$ and consequently 
for the lattice thermal conductivity $\kappa_{ph}$. 

\section{Conclusion}
We presented first principles calculation for the Lorenz function of electron- and 
hole-conducting \SB superlattices and the related bulk materials at varying charge carrier concentration.
As expected, due to bipolar conduction, the Lorenz function increases to large values within 
the intrinsic doping regime. 
More significantly, the Lorenz function L of the superlattices does 
not change monotonically at extrinsic charge 
carrier concentrations. While at increased doping an asymptote convergence of L towards 
the metallic limit $L_0$ is found, a distinct oscillatory behaviour of L is observed. 
This is most pronounced under electron doping and caused by quantum well effects in the 
conduction bands of the superlattices. 
This counterintuitive effect has consequences for 
the determination of the thermal conductivity, as 
L is generally used to separate $\kappa_{el}$ and $\kappa_{ph}$. 
At thermoelectrically profitable charge carrier concentrations the 
application of the metallic value 
$L_0$ to determine the electronic thermal conductivity could lead to a deviation of a factor of two 
in both directions in the worst case. 
Consequently, this leads to wrong estimations of the lattice thermal contribution and the 
figure of merit. A similar behaviour was found theoretically for p-type SiGe superlattices~\cite{ArxivSiGe} 
and this behaviour could be a general effect in thermoelectric superlattices influenced by 
quantum well effects.

\begin{acknowledgements}
This work was supported by the Deutsche For\-schungsgemeinschaft, SPP 1386 `Nanostrukturierte Thermoelektrika: 
Theorie, Modellsysteme und kontrollierte Synthese'. N. F. Hinsche is 
member of the International Max Planck Research School for Science and Technology of Nanostructures. 
\end{acknowledgements}

\bibliography{draft_0.bbl}

\end{document}